\documentclass[sigconf]{acmart}
\usepackage[table]{xcolor}
\usepackage{booktabs,tabularx,array}
\usepackage{placeins}
\usepackage{microtype}
\AtBeginDocument{%
  } 

\newif\ifshowcomments
\showcommentstrue   

\ifshowcomments
  \newcommand{\nasser}[1]{\textcolor{orange!75!black}{\textbf{Nasser:} #1}}
  \newcommand{\valerio}[1]{\textcolor{blue!75!black}{\textbf{Valerio:} #1}}
  \newcommand{\paul}[1]{\textcolor{green!75!black}{\textbf{Paul:} #1}}
  \newcommand{\viraj}[1]{\textcolor{violet!75!black}{\textbf{Viraj:} #1}}
\else
  \newcommand{\nasser}[1]{}
  \newcommand{\valerio}[1]{}
  \newcommand{\paul}[1]{}
  \newcommand{\viraj}[1]{}
\fi

\newlength{\quoteindent}
\colorlet{quotetextcolor}{black!70}
\renewenvironment{quote}
  {\list{}{%
     \leftmargin=\quoteindent
     \rightmargin=\quoteindent
     \topsep=2pt
     \partopsep=0pt
     \parsep=0pt
     \itemsep=0pt}%
   \item\relax\color{quotetextcolor}}
  {\endlist}

\newcommand{\reflectionparagraph}[1]{\par\vskip 2pt\noindent\textbf{#1.}\ }

\renewcommand{\rq}[1]{%
  \ifcase#1\relax
  \or What comments do students write to direct AI code generation, and how do they vary across code constructs?%
  \or To what extent do students modify their comments?%
  \or What are students' perspectives on guiding LLMs to generate code through comments?%
  \else[Undefined research question]%
  \fi 
}

\copyrightyear{2026}
\acmYear{2026}
\setcopyright{cc}
\setcctype{by}
\acmConference[SIGCSE Virtual 2026]{Proceedings of the 2nd ACM Virtual Global Computing Education Conference V.1}{November 12--15, 2026}{Virtual Event, USA}
\acmBooktitle{Proceedings of the 2nd ACM Virtual Global Computing Education Conference V.1 (SIGCSE Virtual 2026), November 12--15, 2026, Virtual Event, USA}
\acmDOI{10.1145/3795867.3830978}
\acmISBN{979-8-4007-2506-7/2026/11}

\begin{document}

\title[Commenting with Copilot]{Commenting with Copilot: A Taxonomy and Multi-Year Analysis of Student Code-Generation Specifications}

\author{Nasser Giacaman}
\affiliation{%
  \institution{University of Auckland}
  \city{Auckland}
  \country{New Zealand}}
\email{n.giacaman@auckland.ac.nz}
\orcid{https://orcid.org/0000-0001-6885-1571}

\author{Valerio Terragni}
\affiliation{%
  \institution{University of Auckland}
  \city{Auckland}
  \country{New Zealand}}
\email{v.terragni@auckland.ac.nz}
\orcid{https://orcid.org/0000-0001-5885-9297}

\author{Paul Denny}
\affiliation{%
  \institution{University of Auckland}
  \city{Auckland}
  \country{New Zealand}}
\email{paul@cs.auckland.ac.nz}
\orcid{https://orcid.org/0000-0002-5150-9806}

\author{Viraj Kumar}
\affiliation{%
  \institution{University of New South Wales}
  \city{Bengaluru}
  \country{India}}
\email{	viraj.kumar1@unsw.edu.au}
\orcid{https://orcid.org/0000-0002-2252-0141}


\begin{abstract}
As AI code tools become integrated into programming environments, students increasingly describe intended behavior in natural language and rely on these tools to generate code, shifting emphasis from code writing to specification. Yet little is known about the comments students write as specifications in AI-assisted programming tasks. We analyze a four-year dataset of undergraduate programming submissions and reflections from tasks in which students wrote comments to guide code generation and refined solutions using test-case feedback. We introduce a taxonomy spanning three dimensions: comment type, 
code expression level, 
and code construct. Using automated classification, we examine how these dimensions vary across attempts and how students describe the process in their reflections. Our findings show that students mostly wrote natural-language \textit{What} comments, shifted toward \textit{How} comments for more procedural constructs, and focused more on verifying generated code than on repeatedly rewriting comments.
\end{abstract}




\maketitle

\section{Introduction}


AI code-generation tools are increasingly integrated into programming environments~\cite{terragni2025future}, changing the skills students need when learning to program~\cite{prather2025beyond}. Rather than writing every line, students can describe intended behavior in natural language and rely on AI tools
to generate candidate implementations~\cite{denny2023conversing}. This has motivated new tasks in computing education. Prompt Problems ask students to solve tasks by writing prompts that elicit correct solutions from a code-generating model~\cite{denny2024prompt}, while dialogue-based prompt programming environments support iterative natural-language interaction with AI models~\cite{padurean2025prompt}. Other work asks students to clarify ambiguous specifications before prompting AI to generate solutions~\cite{pawagi2024probeable}.

However, this skill is not straightforward for novices. Students can struggle to describe their intent, evaluate generated code, and revise prompts when the output is incorrect~\cite{nguyen2024beginning, jiang2022discovering}. Different patterns of AI use may have different implications for learning, with hybrid approaches involving human judgment and verification appearing more promising than simply asking an AI tool for complete solutions~\cite{kazemitabaar2024novices}. There are also concerns that over-reliance on AI might harm core skills such as code reading and comprehension.

In this paper, we analyze student-written comments from four years of an AI-assisted programming activity in an undergraduate object-oriented programming course. Students were shown short Java classes and asked to reproduce their behavior using GitHub Copilot. Rather than writing code themselves, they wrote comments in starter files to guide Copilot, submitted generated code to automated tests, and could revise comments across attempts. 
These comments are not conventional software-engineering comments, but student-written specifications for Copilot. 

We introduce a multidimensional taxonomy that characterizes each comment by its purpose, level of code-like expression, and targeted code construct. 
Our contributions are a taxonomy for analyzing student comments in AI-assisted code generation and a multi-year analysis of how those comments vary across constructs, evolve across attempts, and are described in student reflections. 

The study is guided by three research questions:
 
\begin{description}
  \item[RQ1:] \rq1
    \item[RQ2:] \rq2
    \item[RQ3:] \rq3
\end{description}

\section{Related Work}

\subsection{AI Prompting and Code Generation}

There is broad agreement in computing education that the skills students need are changing, with greater emphasis on reading code and writing natural-language prompts to solve programming tasks~\cite{prather2025beyond}. Prior work has introduced Prompt Problems, in which students solve programming tasks by writing prompts that guide an LLM to generate code~\cite{denny2024prompt}. 
These tasks position prompt writing as practice in expressing computational intent and reflect broader calls to align AI-era learning activities and assessments with intended outcomes, including the intended role of AI~\cite{brabrand2026constructive}. Related work has shown that students craft such prompts in different linguistic forms, including native-language, English, mixed-language, and code-like strategies, highlighting deliberate choices in how programming ideas are expressed in natural language 
\cite{prather2025breaking,smith2024explainplainlanguagequestions}.  

Reading and understanding code remain essential skills, and tasks such as ``Explain in plain English'' have long been used to assess code comprehension~\cite{murphy2012explain, fowler2021how}. Generative AI has recently been leveraged to provide feedback on such tasks by generating code from a student's explanation of a code fragment, thus connecting code comprehension with prompt formulation~\cite{smith2024cgbg,smith2024prompting,denny2024explaining}. 

\subsection{Prompt Revision and Interaction}

AI-assisted programming is often iterative: students write an initial prompt, inspect generated code, test it, and decide whether to revise the prompt, edit the code, or try another strategy. Denny et al. showed that Copilot solved around half of a set of CS1 problems on its first attempt, and that many remaining failures could be resolved through natural-language changes to the problem description~\cite{denny2023conversing}. This demonstrates the value of prompt refinement, although their study used researcher-authored modifications rather than examining how students revise prompts during authentic learning activities. Nguyen et al. directly studied beginning programmers writing and revising prompts for code LLMs, finding that students often struggled to articulate intent, evaluate generated code, and decide how to modify prompts after failures~\cite{nguyen2024beginning}. 

Recent work examines richer forms of iterative interaction. Padu-rean et al. studied dialogue-based prompt programming, where students used multi-turn conversations, code execution, and reflection while solving programming tasks~\cite{padurean2025prompt}. In subsequent work, Padurean et al. showed that students often use prompting to generate an initial solution and then move into short edit--run loops after failed executions, with manual editing increasing with task complexity~\cite{padurean2026interleaving}. These findings align with work on students' trust in Copilot, showing that attitudes toward AI-generated code are shaped by whether students can verify, debug, and understand the output~\cite{shah2025evolution}. Our study builds on this literature by examining a more constrained comment-based workflow, where students do not interact with the model through open dialogue, but revise comments embedded in source files to steer Copilot across attempts.

\subsection{Comments as Specifications}

A further strand of work examines how students use comments to express their understanding of code. In traditional contexts, comments are often treated as documentation for human readers, but they also reveal how students understand program purpose and structure. Kerschbaumer et al. analyzed student comments in a CS1 course and found that high-performing students wrote more task-related explanatory comments, while failing students more often described syntax or implementation details~\cite{kerschbaumer2025comments}. Prior work using the SOLO taxonomy has likewise analyzed students' natural-language explanations of programming processes, showing that such responses vary in completeness and in how they relate programming concepts~\cite{decker2019solo}. Work on subgoal labels also shows that learners can identify functional units in code and express them through labels ranging from line-level descriptions to higher-level functional goals~\cite{jin2024codetree}.

Prior work thus provides a foundation for analyzing comments as expressions of programming intent. In AI-assisted programming, however, comments can also function as prompts or executable specifications, written not only for future human readers but also to influence the code produced by an AI system. This connection is evident in code-generation-based grading and prompting-for-comprehension activities, where students' natural-language explanations of code are used to generate equivalent code and evaluate whether the description captures the intended behavior~\cite{smith2024cgbg,smith2024prompting,denny2024explaining}. Our study extends this line of work by focusing on comments written inside source files to guide Copilot. We examine what comments students write, how these vary across code constructs, whether students revise comments across attempts, and how they perceive the process of guiding LLMs through comments.


\section{Methods}

\subsection{Study Context and Dataset}

This study was conducted in a second-year undergraduate object-oriented programming (OOP) course that used Java to cover classes, objects, encapsulation, interfaces, inheritance, polymorphism, abstract classes, design patterns, exception handling, and basic data structures. 
The Copilot activity ran in the fifth week of the semester, near the course midpoint, after students had covered core object-oriented concepts. 
Students had one week to complete the activity and were not explicitly taught professional prompting strategies for LLM-based code generation or traditional best practices for software comments; instead, the activity focused on using comments as a practical way to express intended behavior for GitHub Copilot.


\subsubsection*{Activity design.}
The handout comprised four tasks: three programming tasks and a final reflection. Students used VS Code with the GitHub Copilot\footnote{Free for students through the GitHub Student Pack \url{https://education.github.com/pack}} extension installed. GitHub's documentation presents two valid ways of obtaining Copilot suggestions\footnote{\url{https://docs.github.com/en/copilot/how-tos/get-code-suggestions/get-ide-code-suggestions?tool=vscode\#getting-code-suggestions-2}}: (i) typing code to initiate inline suggestions, and (ii) describing intent in natural-language comments. We required students to begin with comments rather than handwritten code, both to make them articulate intended behavior before receiving suggestions and to better separate student-authored text from Copilot output.

For each programming task, students were shown a screenshot of a short completed Java class and asked to understand its behavior. The reference solution was provided only as a screenshot so students could inspect the target behavior without copying it into their workspace. They then downloaded starter code with a blank class structure and method stubs and reproduced the behavior by writing only comments to guide Copilot, isolating how they translated that understanding into comment-based guidance for an LLM. 

Students could work incrementally on parts of a class or on the whole class at once. To check progress, they copied the \emph{entire} file, including comments and generated code, into CodeRunner~\cite{coderunner_lobb_2016}, where automated tests showed which behaviors still failed, with no penalty for multiple attempts. Students were explicitly told not to edit the code directly, but to revise their comments in VS Code until the generated code passed the tests. 
Assessment was based on whether their comments led Copilot to generate code that passed the tests, not on exact match to the reference solution.
We assess test-suite adequacy on the reference implementation using two complementary criteria: code coverage with \textsc{JaCoCo}~\cite{jacoco} and mutation score with PIT~\cite{coles2016pit}. Coverage measures the proportion of lines and branch outcomes executed at least once, while mutation score measures the proportion of seeded syntactic faults detected by at least one test~\cite{just2014mutants}.
Per class, the tests achieve: \texttt{SimpleMath} 100\% line, 100\% branch, 82\% mutation;
\texttt{ShoppingCart} 100\% line, 93\% branch, 73\% mutation;
and \texttt{BankAccount} 97\% line, 89\% branch, 84\% mutation.
These values are high, and prior work shows mutation score strongly predicts real-fault detection~\cite{just2014mutants,andrews2005mutation,papadakis2019mutation}. This provides strong evidence that submissions passing the full test suite conform to the specified behavior, though no test suite is exhaustive.


\subsubsection*{Activity tasks.}
The activity included four tasks:
\begin{itemize}
  \item \texttt{SimpleMath}: a small introductory task to help students get comfortable with the workflow; students implemented a simple arithmetic class with basic operations and a counter for negative results.
  \item \texttt{BankAccount}: a more substantial task in which students implemented an account class with balance management, transaction limits, transfers, and formatted account details.
  \item \texttt{ShoppingCart}: a task in which students implemented a shopping cart with item management, total calculation, and a cheapest-item-free discount for larger carts.
  \item Reflection: an open-ended reflection on students' use of Copilot, covering what they found easy or difficult, what they learned, and the tool's benefits, limitations, and dangers; minimum 150 words.
\end{itemize}

\subsubsection*{Dataset construction.}
Our dataset spans four yearly offerings of this activity, from 2023 to 2026. The raw data was exported from CodeRunner as submission logs, including all attempts where students copied their code into CodeRunner and ran the tests. This allowed us to capture the iterative process of writing comments, generating code, checking test feedback, and revising comments.

For analysis, these exports were converted into anonymized per-student folders. Each recorded submission event was reconstructed as an attempt containing the Java file with comments and generated code, the test results, and the reflection response when applicable. This reconstruction preserved the full state of each attempt, which allowed us to study not only the final submitted comments but also how comments changed across repeated submissions. 
 
\subsubsection*{Units of analysis.}
Our main unit of analysis was the student-written comment used to direct code generation. Scripts extracted comments, merged adjacent line comments when they formed a single multi-line comment, and compared successive attempts for each student and question to identify comments as new, unchanged, modified, or deleted.
These extracted comments were then coded using the taxonomy described in Section~\ref{sec:taxonomy}. 

The dataset included 1,161 students and 10,257 recorded submission attempts. Across all three programming tasks, 3,458 of the 3,483 student submissions reached a fully passing solution (99.3\% overall: \texttt{SimpleMath} 100\%, \texttt{BankAccount} 98.8\%, \texttt{ShoppingCart} 98.9\%). Across all attempts, we extracted 136,424 comment instances; counting unchanged carry-forward comments only once reduced this to 52,172 comment instances for content analysis.

\vspace{2mm}

\subsection{Taxonomy}\label{sec:taxonomy}
We coded each extracted student comment on three dimensions: comment type, code expression level, and code construct, assigning one label per dimension. Comments with no meaningful directive, such as empty comments or metadata-only Javadocs, were ignored. 


\subsubsection{Comment Types}
This dimension captures the main role the comment plays as an instruction to the LLM. We adapted the familiar \emph{what/how/why} distinction from prior work on comment purpose~\cite{pascarella2019classifying, cpc_zhai_2020}. 

\emph{What} comments describe the intended result, state, or local behavior that the code should produce. For example:
\begin{quote}
\small\ttfamily\raggedright
// Return false if the amount is less than 0\\
\end{quote}

\emph{How} comments describe a procedure, a sequence of steps, or local control flow. For example:

\begin{quote}
\small\ttfamily\raggedright
// Loop through the arraylist if name is equal to the name of the item in the arraylist then remove it\\
\end{quote}

\emph{Why} comments explain the reason for an action or constraint. For example:
\begin{quote}
\small\ttfamily\raggedright
// Not returning the cheapest price as we need to find the cheapest price in the cart first\\
\end{quote}

\subsubsection{Code Expression Level}

This dimension captures how closely a comment resembles executable or near-executable code.

\emph{Strict-Code} comments are essentially code or pseudocode that is \emph{almost directly compilable}, requiring at most trivial surface changes such as adding a semicolon. For example:
\begin{quote}
\small\ttfamily\raggedright
// balance = balance + amount\\
\end{quote}

\emph{Code-Like} comments are written in words and very close to the local implementation, but unlike \emph{Strict-Code}, would \emph{not compile as are}; instead, they read like direct paraphrases of code. For example:

\begin{quote}
\small\ttfamily\raggedright
// set balance to balance + amount\\
\end{quote}

\emph{Natural-Language} comments are ordinary instructions or explanations that do not closely resemble code. For example:
\begin{quote}
\small\ttfamily\raggedright
// If more than 5 items are bought, discount cheapest one\\
\end{quote}

\vspace{4mm}

\begin{table*}[!t]
\centering
\caption{Comment type and expression level by code construct (ignoring unmodified comments across attempts)}
\label{tab:rq1-construct-summary}
\resizebox{0.85\textwidth}{!}{%
\begin{tabular}{lrr rrr rrr} 
\hline
 & & \multicolumn{3}{c}{\textbf{Comment Type}} & \multicolumn{3}{c}{\textbf{Expression Level}} \\
\cline{3-5} \cline{6-8}
\textbf{Construct} 
& \textbf{Count} 
& \textbf{What} 
& \textbf{How} 
& \textbf{Why} 
& \textbf{Natural Lang} 
& \textbf{Code-like} 
& \textbf{Strict Code} \\
\hline
Multi-Step-Block & 11,067 (16.3\%) & 617 (5.6\%) & \textbf{10,431 (94.3\%)} & 19 (0.2\%) & \textbf{10,854 (98.1\%)} & 166 (1.5\%) & 47 (0.4\%) \\
Conditional-Return & 8,879 (13.1\%) & \textbf{8,529 (96.1\%)} & 260 (2.9\%) & 90 (1.0\%) & \textbf{8,745 (98.5\%)} & 88 (1.0\%) & 46 (0.5\%) \\
Arithmetic-Update & 8,941 (13.2\%) & \textbf{8,564 (95.8\%)} & 343 (3.8\%) & 34 (0.4\%) & \textbf{8,310 (92.9\%)} & 496 (5.5\%) & 135 (1.5\%) \\
Conditional & 8,739 (12.9\%) & \textbf{7,075 (81.0\%)} & 1,601 (18.3\%) & 63 (0.7\%) & \textbf{8,627 (98.7\%)} & 89 (1.0\%) & 23 (0.3\%) \\
Return & 9,169 (13.5\%) & \textbf{9,131 (99.6\%)} & 15 (0.2\%) & 23 (0.3\%) & \textbf{5,714 (62.3\%)} & 94 (1.0\%) & 3,361 (36.7\%) \\
Assignment-Init. & 6,889 (10.2\%) & \textbf{6,618 (96.1\%)} & 261 (3.8\%) & 10 (0.1\%) & \textbf{6,643 (96.4\%)} & 201 (2.9\%) & 45 (0.7\%) \\
Field-Declaration & 4,393 (6.5\%)& \textbf{4,372 (99.5\%)} & 4 (0.1\%) & 17 (0.4\%) & \textbf{4,274 (97.3\%)} & 85 (1.9\%) & 34 (0.8\%) \\
Output-Formatting & 2,803 (4.1\%)& \textbf{2,505 (89.4\%)} & 295 (10.5\%) & 3 (0.1\%) & \textbf{2,602 (92.8\%)} & 174 (6.2\%) & 27 (1.0\%) \\
Method-Invocation & 2,171 (3.2\%)& \textbf{2,097 (96.6\%)} & 62 (2.9\%) & 12 (0.6\%) & \textbf{2,115 (97.4\%)} & 30 (1.4\%) & 26 (1.2\%) \\
Loop-Iteration & 1,657 (2.4\%)& 707 (42.7\%) & \textbf{949 (57.3\%)} & 1 (0.1\%) & \textbf{1,577 (95.2\%)} & 52 (3.1\%) & 28 (1.7\%) \\
Other & 856 (1.3\%)& \textbf{822 (96.0\%)} & 9 (1.1\%) & 25 (2.9\%) & \textbf{839 (98.0\%)} & 14 (1.6\%) & 3 (0.4\%) \\
Object-Creation & 704 (1.0\%) & \textbf{702 (99.7\%)} & 2 (0.3\%) & 0 (0.0\%) & \textbf{684 (97.2\%)} & 12 (1.7\%) & 8 (1.1\%) \\
Local-Declaration & 434 (0.6\%)& \textbf{433 (99.8\%)} & 1 (0.2\%) & 0 (0.0\%) & \textbf{419 (96.5\%)} & 6 (1.4\%) & 9 (2.1\%) \\
Ignored & 1,085 (1.6 \%)& -  & - &  - & -  & - & - \\
\hline
Overall & 67,787 (100.0\%) & \textbf{52,172 (77.0\%)} & 14,233 (21.0\%) & 297 (0.4\%) & \textbf{61,403 (90.6\%)} & 1,507 (2.2\%) & 3,792 (5.6\%) \\
\hline
\end{tabular}
}
\end{table*}

\subsubsection{Code Constructs}

This dimension captures the program element targeted by a comment. The final set combined Java constructs with a few study-specific aggregations that better matched the task structure and the instructions students wrote:

\begin{enumerate}
  \item \textbf{Field-Declaration}: declares a class-level member variable.
  \item \textbf{Local-Declaration}: declares a local variable in a method.
  \item \textbf{Assignment-Initialization}: assigns or initializes a value, including in a declaration.
  \item \textbf{Conditional}: an \texttt{if} or \texttt{if-else} branch.
  \item \textbf{Loop-Iteration}: repeated traversal using a \texttt{for}/\texttt{while} loop.
  \item \textbf{Object-Creation}: creates a new object, typically with \texttt{new}.
  \item \textbf{Method-Invocation}: calls an existing method.
  \item \textbf{Return}: returns a value directly, without the return being defined by a condition.
  \item \textbf{Conditional-Return}: returns a value because a condition is met, e.g. \texttt{if (...) return false}.
  \item \textbf{Arithmetic-Update}: changes a numeric value.
  \item \textbf{Output-Formatting}: prints, displays, or formats output.
  \item \textbf{Multi-Step-Block}: used when multiple actions or constructs are covered and no one category dominates.
  \item \textbf{Other}: fallback when none of the above fits clearly.
\end{enumerate}


\subsection{LLM Classification Pipelines}

Recent work has explored LLM-supported labeling workflows, including deductive content analysis, collaborative codebook refinement, multi-agent approaches, and researcher-controlled coding~\cite{chew2023llm,dai2023llm,qiao2025thematic,sharma2025details}. We therefore used GPT-5.4-family models with medium reasoning in two pipelines: one to label extracted comments, and one to organize student reflections while preserving traceability through supporting quotations and manual verification.

\paragraph{Comment classification.}
Each extracted comment was classified with an LLM using the full Java class from the corresponding attempt as context, assigning the comment type, expression level, and code construct labels described in Section~\ref{sec:taxonomy}. The labeled dataset was then used for the quantitative analyses reported in Section~\ref{sec:results}.

To assess the reliability of the LLM classification, we randomly sampled 200 extracted comments. One researcher independently labelled the sample using the taxonomy described in Section~\ref{sec:taxonomy} while blinded to the LLM-generated labels. The human and LLM classifications showed high agreement across all three taxonomy dimensions. For Comment Type, agreement was 92.5\% ($\kappa=0.781$); for Expression Level, agreement was 94.0\% ($\kappa=0.771$); and for Code Construct, agreement was 93.5\% ($\kappa=0.927$).


\paragraph{Reflection classification.}
We analyzed the end-of-activity reflections with an LLM-assisted four-round pipeline. 
Reflection responses were exported from CodeRunner, assigned coded student identifiers, and analyzed in four rounds: (i) generating candidate codes for individual reflections, (ii) consolidating them into a shared code set, (iii) applying that code set back to the full set of reflections, and (iv) grouping the final codes into broader themes.


\section{Results} \label{sec:results}

The following results aggregate data across all four yearly offerings of the activity. As patterns were broadly similar across cohorts, results are presented in aggregate for clarity.

\subsection{RQ1: Comment Types and Expression}

Table~\ref{tab:rq1-construct-summary} summarizes the distribution of comment types and expression levels across code constructs. Overall, comments were dominated by \textit{What} comments (77.0\%), followed by \textit{How} comments (21.0\%), while \textit{Why} comments were rare (0.4\%). Most constructs were likewise dominated by \textit{What} comments. The exceptions were \textit{Multi-Step-Block}, where 94.3\% of comments were classified as \textit{How}, and \textit{Loop-Iteration}, where 57.3\% were classified as \textit{How}. This suggests that when logic involved multiple steps or repeated actions, students shifted from describing intended outcomes to describing the procedure needed to produce them. Comments were predominantly expressed in natural language, except for \textit{Return} statements, which showed a fairly high proportion of \textit{Strict-Code} expressions.


\subsection{RQ2: Extent of Comment Modification}

\begin{table}[!tbp]
\centering
\caption{Themes and code labels from student reflections}
\label{tab:abridged-theme-code-labels}
\scriptsize
\setlength{\tabcolsep}{3pt}
\renewcommand{\arraystretch}{0.96}
\begin{tabularx}{\columnwidth}{@{}>{\centering\arraybackslash}p{0.05\columnwidth} >{\raggedright\arraybackslash}p{0.12\columnwidth} >{\centering\arraybackslash}p{0.1\columnwidth} X@{}}
\toprule
\textbf{Rank} & \textbf{Theme} & \textbf{Students} & \textbf{Code labels} \\
\midrule
1 & Speed and scaffolding & \parbox[t]{\linewidth}{\centering 84.5\%\\979/1158} & Faster drafting and coding~(830), Autocomplete reduces writing effort~(331), Intent to implementation details~(198), Helps students get started~(125),  Time pressure boosts AI appeal~(9) \\
\addlinespace[5pt]
2 & Task fit & \parbox[t]{\linewidth}{\centering 64.5\%\\747/1158} & Best for boilerplate code~(301), Best on familiar simple tasks~(236),  Struggles on complex specific tasks~(219), Benefits depend on user readiness~(136),  Prompting slower for small tasks~(135), Best when solution is known~(125),  Weaker on large codebases~(40), Learned patterns shape quality~(28) \\
\addlinespace[5pt]
3 & Verification burden & \parbox[t]{\linewidth}{\centering 57.3\%\\663/1158} & Understand review and validate output~(444), AI still needs oversight~(418),  Checking can erode time savings~(99), Limited trust without understanding~(77),  Test against expected behavior~(53), Easy suggestions invite overtrust~(51), Can introduce errors~(29) \\
\addlinespace[5pt]
4 & Learning and agency & \parbox[t]{\linewidth}{\centering 57.2\%\\662/1158} & Overreliance can weaken learning~(576), Understanding over AI output~(170),  AI-supported practice can teach~(68), Can reduce agency and satisfaction~(36) \\
\addlinespace[5pt]
5 & Code quality & \parbox[t]{\linewidth}{\centering 54.5\%\\631/1158} & Output may be incomplete/wrong~(409), May add unrequested code~(99),  Output often needs cleanup~(92), Works but may be inefficient~(91),  Misses exact technical details~(91), Plausible code can hide errors~(73),  Unfamiliar style hinders maintenance~(25), Long outputs resist inspection~(19),  Mixed help on declarations details~(11), May improve code consistency~(10) \\
\addlinespace[5pt] 
6 & Prompt precision & \parbox[t]{\linewidth}{\centering 54.0\%\\625/1158} & Clear specific prompts help~(405), State specifics explicitly~(196),  Vague prompts derail output~(119), Right wording is tedious~(85),  Exact output details are hard~(70), Right terms express intent~(19) \\
\addlinespace[5pt]
7 & Workflow adoption & \parbox[t]{\linewidth}{\centering 53.3\%\\617/1158} & Mixed but pragmatic value~(567), Setup, responsiveness matter~(60),  Positive overall, few limits~(29), Convenient editor use~(15),  AI use may keep growing~(7), Compared with other tools~(3) \\
\addlinespace[5pt]
8 & Comment interface & \parbox[t]{\linewidth}{\centering 51.6\%\\598/1158} & Easy comment-based prompting~(474), Comments express intended behavior~(156),  Autocompletes comments and prompts~(90), Comments also aid understanding~(39) \\
\addlinespace[5pt]
9 & Context grounding & \parbox[t]{\linewidth}{\centering 37.5\%\\434/1158} & Comments may miss intent~(254), Nearby context steers suggestions~(170),  Missing context hurts quality~(68) \\
\addlinespace[5pt]
10 & Selective partnership & \parbox[t]{\linewidth}{\centering 36.5\%\\423/1158} & Support tool, not replacement~(285), Suggests simpler better methods~(118),  Suggestions shape next steps~(38), Adapt and refine outputs~(33),  Scaffolds debugging when stuck~(23) \\
\addlinespace[5pt]
11 & Steering work & \parbox[t]{\linewidth}{\centering 35.5\%\\411/1158} & Iteratively refine prompts~(218), Prompting improves with practice~(92),  Wrong paths can persist~(66), Planning and code reading help~(50),  Break hard tasks into steps~(44), Failures are hard to diagnose~(14),  Wording shifts can be unpredictable~(9) \\
\addlinespace[5pt]
12 & Predictive leaps & \parbox[t]{\linewidth}{\centering 27.7\%\\321/1158} & Impressively accurate~(133), High-level intent, AI fills details~(97),  Small cues trigger large completions~(66), Short cues work when obvious~(57),  Impressively capable, sometimes uncanny~(50) \\
\addlinespace[5pt]
13 & Ethical risks & \parbox[t]{\linewidth}{\centering 17.9\%\\207/1158} & Authorship, plagiarism, security concerns~(122), Job replacement worries~(61),  Privacy and confidentiality concerns~(23), Broader ethical, safety, social concerns~(20),  May enable harmful misuse~(3) \\
\bottomrule
\end{tabularx}
\end{table}

\begin{figure*}[!t]
\centering
\includegraphics[width=\textwidth]{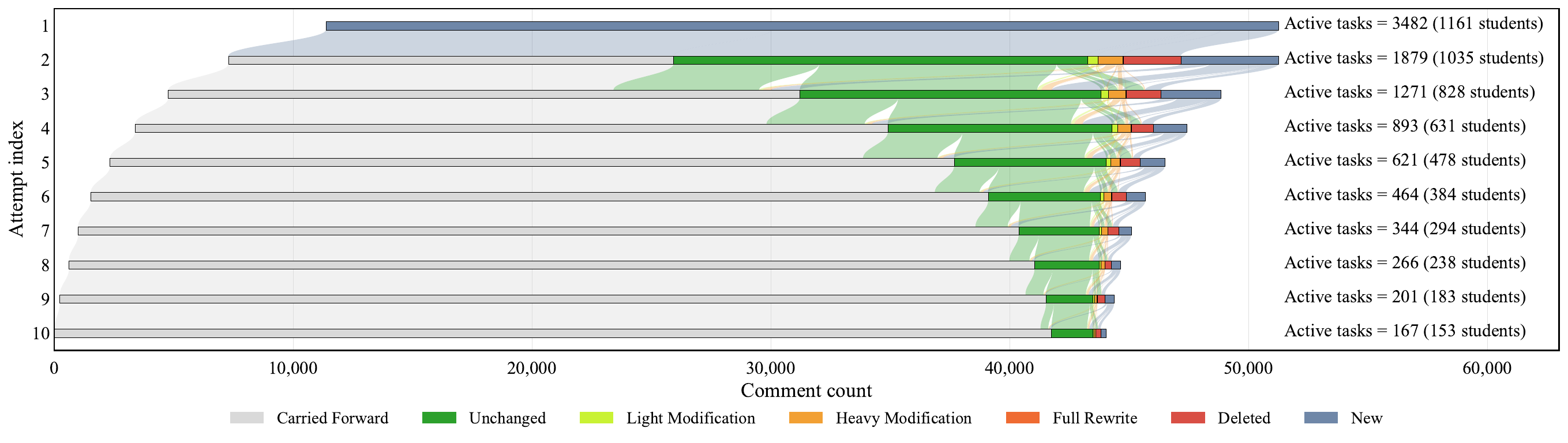}
\caption{Pooled comment modification categories across all three exercises and four yearly cohorts for the first 10 attempts of each student--task sequence; the grey \textit{Carried Forward} segment marks comments retained in the concluded attempt.}
\label{fig:comment-modification-all-years}
\end{figure*}

Figure~\ref{fig:comment-modification-all-years} summarizes comment modifications across student attempts. The grey \textit{Carried Forward} segment marks comments still present in the student's concluded attempt, rather than comments unchanged in still-active attempts. The dominance of \textit{Unchanged} and \textit{Carried Forward} suggests that students usually kept comments rather than substantially rewriting them. The figure shows only the first 10 attempts, covering 124,201 extracted comments (91.0\% overall) and 51,248 unchanged comments (98.2\% of all unchanged comments), indicating that most comment writing and modification occurred early. Within this window, \textit{Unchanged} (48.4\%) and \textit{New} (41.3\%) dominate, followed by \textit{Deleted} (6.0\%); \textit{Light Modification}, \textit{Heavy Modification}, and \textit{Full Rewrite} together account for only 4.4\%.

These modification levels were assigned automatically using Levenshtein similarity between successive comments. Thresholds were selected following inspection of representative comment pairs to distinguish minor wording edits from substantial revisions. Similarity scores of at least 0.8 consistently reflected minor wording changes and were classified as \textit{Light Modification}. Scores below 0.2 generally corresponded to comments that had been rewritten with substantially different wording or intent and were classified as \textit{Full Rewrite}. Intermediate values were classified as \textit{Heavy Modification}.

\begin{figure*}[!t]
\centering
\vspace{-1mm}
\includegraphics[width=0.86\textwidth]{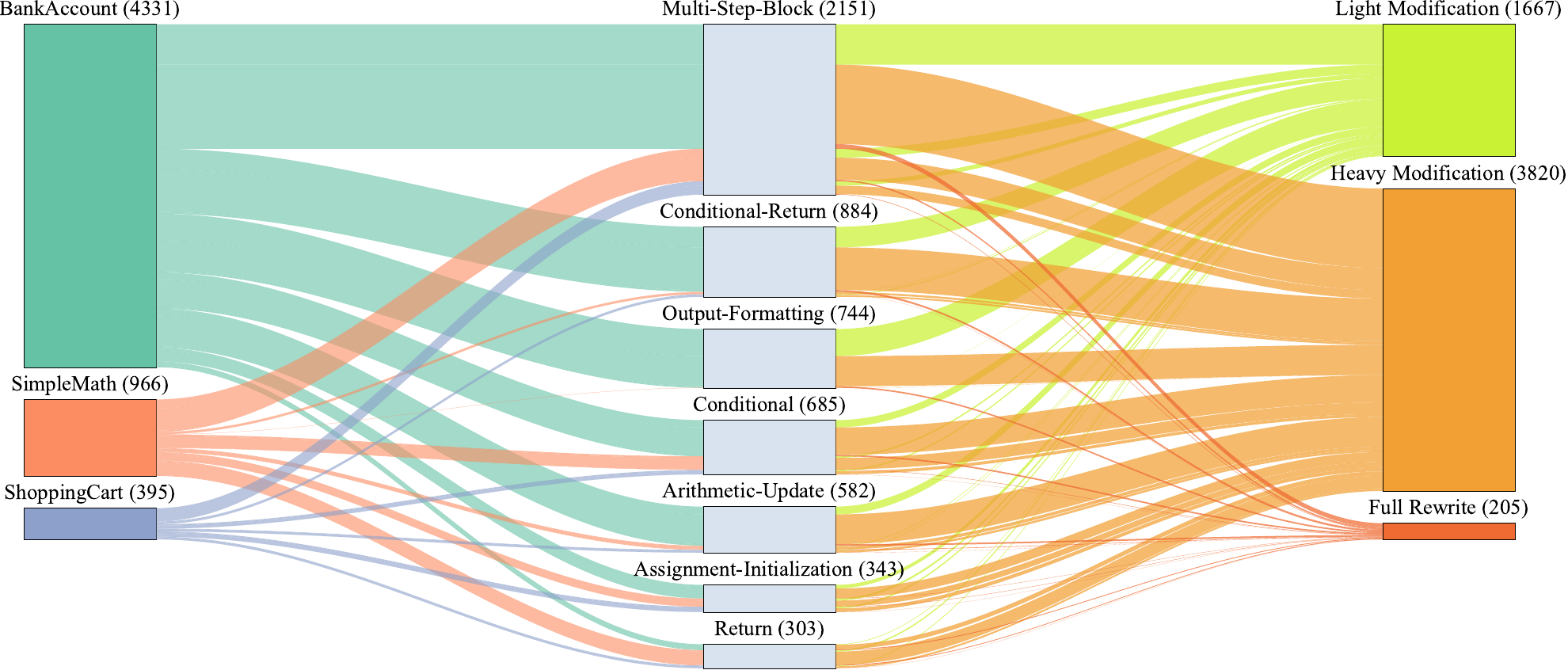}
\vspace{-2mm}
\caption{Changed comments from question to construct to modification category. The top seven constructs are shown (5,692 of 6,221 events); omitted constructs account for 529 events (8.5\%).}
\label{fig:question-construct-revision-effort}
\end{figure*}

To examine where revision effort was concentrated, Figure~\ref{fig:question-construct-revision-effort} traces changes from question to construct to modification category for the top seven constructs. \textit{BankAccount} dominates with the most change events. \textit{Multi-Step-Block} is the clearest hotspot, followed by \textit{Conditional-Return}, \textit{Output-Formatting}, \textit{Conditional}, and \textit{Arithmetic-Update}. The widest flows end in \textit{Heavy Modification}, especially for \textit{Multi-Step-Block}, suggesting students more often needed substantial restructuring than minor edits. 

\subsection{RQ3: Student Perspectives}

Table~\ref{tab:abridged-theme-code-labels} summarizes 13 themes; the six most prevalent are discussed.

\reflectionparagraph{Speed and scaffolding}
This was the dominant theme. Students often described comments as a quick way to turn ideas into code, especially when getting started or avoiding low-level syntax: 
 
\begin{quote}
\small\itshape\raggedright
 ``It speeds up development by suggesting code in real time, reducing the need to manually write repetitive sections or search for solutions.''
\end{quote}

\reflectionparagraph{Task fit}
Copilot was not described as equally useful across tasks. Instead, they saw it as strongest on familiar or repetitive work, and weaker for niche, multi-step, or constrained requirements: 

\begin{quote}
\small\itshape\raggedright
``Copilot tends to give solutions that are generic or repeated, and it can struggle to complete code that has to be done a specific way.''
\end{quote}

\reflectionparagraph{Verification burden}
Even when students valued the tool, they framed checking and interpreting output as their responsibility. This helps explain the RQ2 revision results: the low rate of explicit comment rewriting does not imply a smooth process: 

\begin{quote}
\small\itshape\raggedright
``I had to keep checking the output and fix mistakes, especially when the logic was slightly wrong or when it misunderstood the instructions.''
\end{quote}

\reflectionparagraph{Learning and agency}
Some students described learning gains when using the output critically, while many worried that relying too heavily on Copilot would negatively impact learning: 

\begin{quote}
\small\itshape\raggedright
``Heavily relying on Copilot as a crutch for programming limits your own ability to gain programming knowledge effectively and can be detrimental if you ever find yourself with a problem that generative AI cannot solve.''
\end{quote}

\reflectionparagraph{Code quality}
Students often noted that Copilot's output could look plausible yet be wrong or introduce unwanted behavior: 

\begin{quote}
\small\itshape\raggedright
``Sometimes it tells you the wrong code and insists that it is correct and doesn't offer other options despite changing the description.''
\end{quote}

\reflectionparagraph{Prompt precision}
Students repeatedly reported that they had to spell out constraints, sequencing, and exact method choices if they wanted reliable results, especially on the more complex tasks: 

\begin{quote}
\small\itshape\raggedright
``I learnt that you have to be very direct and specific of what you want the AI to do so it avoids going on tangent.''
\end{quote}


\section{Discussion}

The results suggest that students usually approached comment-based code generation by stating intended behavior in plain language rather than writing code-like prompts.
Most comments were \textit{What} comments, especially for simpler constructs, while more procedural constructs 
drew more \textit{How} comments. An interpretation is that students often described local outcomes when the behavior was easy to express, but shifted toward stepped guidance when the logic required sequencing or repetition.

The modification results support this picture, as most comments were either new or unchanged with relatively little rewriting. The reflections suggest that this does not mean the process was easy. Instead, many students described the main effort as checking generated code, interpreting test results, and deciding whether output matched intended behavior. This explains why comment revision was limited even though verification burden was a key theme; work shifted from rewriting prompts to reviewing generated code.


For teaching, these findings suggest that comment-based AI activities may be most useful when framed as exercises in understanding, expressing, and verifying behavior rather than simply obtaining code~\cite{denny2024explaining}. Students often described Copilot as most useful for familiar tasks, but less reliable for multi-step or tightly constrained ones. This matches the greater revision effort on procedural constructs.
This supports careful AI use in which students remain responsible for interpreting and checking output, rather than treating the tool as an unquestioned answer source~\cite{shah2025evolution}.

This study has several limitations. It comes from one course, one institution, and one activity design, so the results may not transfer directly to other settings. 
Students reproduced short Java examples from solution screenshots rather than solving open-ended problems, so the findings reflect a constrained programming task.
The modification analysis captures only changes visible in submissions; IDE-only revisions overwritten before submission do not appear in our data. Passing the tests also does not guarantee complete semantic equivalence to the reference solutions. Finally, both the comment classification and reflection analysis relied on LLM-assisted pipelines, so classification errors exist.

\section{Conclusions}

This paper examined how students wrote comments as specifications in a four-year GitHub Copilot activity. Students primarily wrote natural-language \textit{What} comments, shifted toward more procedural \textit{How} comments for multi-step and iterative constructs, and revised comments less often than they reviewed and validated generated code. These results suggest that comment-based AI programming is not simply a matter of asking for code: it is a combined task of understanding behavior, expressing intent clearly, and judging whether the generated solution is correct. For computing education, this points toward treating specification and verification as central learning goals in AI-assisted programming, rather than framing AI tools primarily as shortcuts for code production.



\bibliographystyle{ACM-Reference-Format}
\bibliography{references}


\end{document}